\documentclass[12pt,dvips]{article}

\usepackage{amssymb} 
\usepackage{epsfig}

\newcommand{\RE}{\mathbb{R}}
\newcommand{\be}{\begin{equation}}
\newcommand{\ee}{\end{equation}}
\newcommand{\bea}{\begin{eqnarray}}
\newcommand{\eea}{\end{eqnarray}}
\newcommand{\f}{\frac}
\newcommand{\R}{{\rm Re}}
\newcommand{\I}{{\rm Im}}
\newcommand{\subcond}{\ensuremath{\langle\bar{\psi}_x\psi_x\rangle_{\rm sub}}{}}
\newcommand\M{\ensuremath{{\cal M\,}}{}}
\newcommand\Z{\ensuremath{{\cal Z\,}}{}}
\newcommand\D{\ensuremath{{\cal D\,}}{}}

\begin{document}

\thispagestyle{empty}

\date{February 23, 1999}
\title{
\vspace{-5.0cm}
\begin{flushright}
{\normalsize UNIGRAZ-UTP-23-2-99}\\
\end{flushright}
\vspace*{2cm}
Leutwyler--Smilga  sum rules for Ginsparg--Wilson 
lattice fermions\thanks{Supported by Fonds 
zur F\"orderung der Wissenschaftlichen Forschung 
in \"Osterreich, Project P11502-PHY.} }
\author
{\bf F. Farchioni  \\
Institut f\"ur Theoretische Physik,\\
Universit\"at Graz, A-8010 Graz, AUSTRIA}
\maketitle
\begin{abstract}
We argue that lattice QCD with Ginsparg--Wilson fermions
satisfies the Leutwyler--Smilga sum rules for 
the eigenvalues of the chiral Dirac operator.
The result is obtained in the one flavor case, by rephrasing 
Leutwyler and Smilga's original analysis 
for the finite volume partition function.
This is a further evidence that Ginsparg--Wilson fermions,
even if breaking explicitly the chirality on the lattice
in accordance to the Nielsen--Ninomiya theorem, mimic 
the main features of the continuum theory related
to chiral symmetry.

\end{abstract}

\newpage

\section{Introduction}
\label{sec:intro}

According to the Nielsen--Ninomiya theorem \cite{NiNi81a} there is no way to
avoid the explicit breaking on the lattice of the chiral symmetry of the
continuum QCD (at least if one wants to keep the fundamental
locality property). 
In the simplest discretization, given by the Wilson action, the explicit 
breaking  introduces many annoying artifacts in the lattice theory, 
the most popular being the quark mass renormalization
with related fine tuning problem; more in general, features and 
mechanisms of the continuum theory associated to the chirality find 
no correspondence in the lattice theory.
For example, the Atiyah--Singer theorem \cite{AtSi71}, 
relating the chirality of the 
zero modes of the Dirac operator in a finite volume to the 
topological charge of the background configuration, has no lattice
counterpart; the study of the spontaneous breaking of the symmetry
is awkward since the explicit breaking introduces spurious 
effects which are not under full theoretical control 
(e.g. an order parameter analogous to the fermion condensate is missing).

Ginsparg and Wilson provided the condition \cite{GiWi82} under which
the breaking of the chirality of the lattice action is the mildest 
possible, compatibly with the Nielsen--Ninomiya theorem.
This is the so called Ginsparg--Wilson condition (GWC)
for the lattice Dirac operator.

Recently, it has been realized that the Dirac operator associated to 
a fixed point of a renormalization group transformation \cite{Ha98c}, and the
one coming from the `overlap formalism' \cite{ovlap}, both satisfy the GWC
\cite{Ha98c,Ne98a}.
Further analysis has shown that the main mechanisms of
QCD related to chiral symmetry find correspondence on
the lattice with Ginsparg--Wilson fermions \cite{HaLaNi98,Ha98a,Lu98-Ch98}. 
Formulation of chiral gauge theories describing Weyl fermions
is also possible \cite{Lu98a}.
All this is somehow miraculous, taking place in a scenario 
where the chirality {\em is explicitly broken}.

Here we concentrate on one particular aspect, i.e. the 
Leutwyler--Smilga sum rules \cite{LeSm92} 
for the eigenvalues of the Dirac
operator of chiral QCD. These can be explicitly derived in the continuum
(regularized) theory, from the analysis of the finite volume partition 
function in a limit (large volumes and small quark masses) where QCD reduces 
to a simple matrix model and chiral symmetry plays the fundamental role.
Starting from this, Shuryak and Verbaarschot put forward the hypothesis
\cite{ShVe93} that the sum rules are an universal feature, 
i.e. model independent, 
the only precondition being chiral symmetry. The simplest chiral model 
is a chiral Random Matrix Theory \cite{ShVe93}; 
here Leutwyler--Smilga sum rules 
have been verified, enabling also a systematic classification of the 
different universal behaviors \cite{Ve94}.
First checks of the predictions from chiral Random Matrix Theory 
have been done in the framework of
lattice gauge theory  with staggered fermions \cite{rmt-latt}, 
which however restore the chirality on the lattice only partially.

Ginsparg--Wilson fermions, which {\em effectively} restore chiral symmetry,
appear the ideal environment for studying \cite{rmt-gw} Shuryak 
and Verbaarschot's chirality--induced universality on the lattice. 
Equivalence of a Random Matrix Theory complying with 
the GWC to the ordinary chiral Random Matrix Theory
has been proved in \cite{Sp98}.

Here, we explicitly show how Leutwyler--Smilga sum rules are 
recovered in lattice QCD, in a framework of explicitly broken chiral symmetry,
with Ginsparg--Wilson fermions.\footnote{After the completion of the work, 
the author realized that this result was already derived, with a 
slightly different approach, by F. Niedermayer in his talk \cite{Ni98} 
at Lattice 98 conference.}We consider 
the simplest case of just one quark flavor, 
where the absence of massless excitations allows to avoid 
the complications related to the management of finite--size effects.
We follow the line of reasoning of the seminal paper 
\cite{LeSm92}.

\section{Finite volume partition function}
\label{sec:part}

We start from the lattice action
\be
S\:=\:S_G(U) \,+\, \sum_x\:\bar{\psi}_x(\D_{x,x^{\prime}}(U)+\M)\psi_{x'}\;\;,
\ee
(Dirac and color indices are omitted) where 
\be 
\M\:=\: m\, \f{(1+\gamma^5)}{2}\, +\, m^*\,\f{(1-\gamma^5)}{2}\;\;
\ee
(m is a complex mass); $\D$ is any Ginsparg--Wilson Dirac operator
satisfying
\be
\left\{\, \D_{x,x^{\prime}},\gamma^5\,\right\}\;=\;\left(\,
\D\,\gamma^5\,\D\,\right)_{x,x^{\prime}}\;\;.
\label{eq:gwc}
\ee
The latter relation is a particular case of a broader (but still equivalent)
condition
\be
\left\{\, \D_{x,x^{\prime}},\gamma^5\,\right\}\;=\;\left(\,
2\,\D\,\gamma^5\,R\,\D\,\right)_{x,x^{\prime}}\;\;,
\label{eq:gwcg}
\ee
where $R$ is an operator depending only on color and space--time (not Dirac)
indices, whose matrix elements have finite range 
in space--time indices, or exponentially 
decay with the distance. Our case corresponds to 
$R_{x,x'}=(1/2)\,\delta_{x,x'}$.
We assume for $\D$ also the $\gamma^5$--hermiticity property
$\D^{\dagger}=\gamma^5\D\gamma^5$ which implies in particular
that the spectrum is invariant under complex conjugation
$\lambda\rightarrow\overline{\lambda}$.
We consider the theory on a finite lattice of extension
$L$, with periodic (anti--periodic) boundary conditions for
gauge (fermionic) degrees of freedom.
All quantities are expressed in lattice units.

As a consequence of (\ref{eq:gwc}) the spectrum of $\D$  $\{\lambda\}$
is constrained on a unit circle in the complex plane centered in (1,0) 
\be
\label{eq:constr}
|\lambda-1|^2\:=\:1\;\;; 
\ee
the set of its eigenvectors $\{v_\lambda\}$ form
an orthonormal complete basis, and have definite chiral 
properties:
\be
\label{eq:chipr}
\gamma^5\,v_{\lambda}\:=\:\left\{
\begin{array}{ll}
v_{\overline{\lambda}} & \;\;\;\mbox{if $\lambda\not=\overline{\lambda}$}\\
\pm\,v_{\lambda} & \;\;\;\mbox{if $\lambda\in\RE$}\;\;.
\end{array}
\right.
\ee
In particular, the index of $\D(U)$, $\nu(U)$, can be 
defined \cite{HaLaNi98} as the number of zero modes counted with their 
chirality. \footnote{We assume \cite{LeSm92} that different 
chiralities cannot mix.}

In analogy with the continuum, the partition function $\Z(\theta,m)$ is
defined as:
\be
\Z(\theta,m)\:=\:\sum_{\nu=-\infty}^{\infty}\:e^{i\theta\nu}\,\Z_{\nu}(m)\;\;,
\ee
where $\Z_{\nu}(m)$ is obtained by integrating just on configurations with
associated index $\nu$:
\be
\label{eq:znu}
\Z_{\nu}(m)\:=\:\int[dU^{(\nu)}] \,e^{-\beta S_G}\,{\rm det}
(\D(U^{(\nu)})+\M)\;\;.
\ee
As a consequence of (\ref{eq:chipr}) the matrix $\M$ is (as in the continuum)
block--diagonal in the basis $\{v_\lambda\}$, each block living in 
the 2--dimensional subspace spanned by $v_{\lambda}$ and 
$v_{\overline{\lambda}}\:$:
\be
\left(
\begin{array}{cc}
\R(m) &  i\,\I(m) \\ 
i\,\I(m) & \R(m)
\end{array}
\right)\;\;;
\ee
in the case $\lambda\in{\rm \RE}$ (i.e., because of (\ref{eq:constr}), 
$\lambda=0,2$) $v_{\lambda}$ is also 
an eigenvector of $\M$ with eigenvalue $m$ for positive chirality  
and $m^*$ for negative chirality.

Using these properties and exploiting the constraint (\ref{eq:constr})
as well, we can write an explicit expression for the fermion determinant 
in (\ref{eq:znu}) in terms of the eigenvalues of $\D$; for $\nu>0$:
\bea
\label{eq:det}
{\rm det}(\D(U^{(\nu)})+\M)\:=\:(2+m)^{N^+_{rm}}\,(2+m^*)^{N^-_{rm}}\nonumber\\
{m^{\nu}\,\prod_{i}}''\,\left(\,(1+\R(m))\,|\lambda_i|^2+|m|^2\,\right)\;\;,
\eea
where the double--primed product indicates the product over half of 
the complex eigenvalues and $N^+_{rm}$ ($N^-_{rm}$) is the number 
of positive (negative) real modes;
for $\nu<0$, $m^{\nu}\rightarrow (m^*)^{-\nu}$ 
in the above formula.

Following \cite{LeSm92} we now consider the problem 
of finding  an explicit representation for $\Z(\theta,m)$ in the limit
\be
\label{eq:lim}
L\rightarrow\infty, \;\;\; m\rightarrow 0, \;\;\; m\,L^d=const\;\;
\ee
($d$ is the number of space--time dimensions).
We rewrite the determinant in (\ref{eq:det}):
\bea
{\rm det}(\D(U^{(\nu)})+\M)\:=\:(2+m)^{N^+_{rm}}\,
(2+m^*)^{N^-_{rm}}\;\;\;\;\;\;\;\;\;\;\;\nonumber\\
\left[(1+\R(m))^{\f{1}{2}\,(N_c\,N_D\,L^d-\nu-N_{rm})}\right]\,m^{\nu}\,{\prod_{i}}''\,
\left(\,|\lambda_i|^2+\f{|m|^2}{(1+\R(m))}\,\right)\;\;,
\eea
where $N_c$ and $N_D$ denote the number of color and Dirac degrees of freedom
respectively and $N_{rm}=N^+_{rm}+N^-_{rm}$.
In the limit (\ref{eq:lim}) we can replace
\bea 
(1+\R(m))^{\f{1}{2}\,(N_c\,N_D\,L^d-\nu-N_{rm})}&\:\rightarrow\:&e^{\f{1}{2}N_c\,N_D\,L^d\,\R(m)}\;\;,\nonumber\\
\f{|m|^2}{(1+\R(m))}&\:\rightarrow\:&|m|^2
\eea
and the factor related to the real modes reduces to $2^{N_{rm}}$;
we obtain
\be
{\rm det}(\D(U^{(\nu)})+\M)\:\rightarrow\:
\,
e^{\f{1}{2}N_c\,N_D\,L^d\,\R(m)}\,2^{N_{rm}}\,m^{\nu}\,{\prod_{i}}''\,\left(\,|\lambda_i|^2+|m|^2\,\right)\;\;.
\ee
Apart from the multiplicative factor $e^{\f{1}{2}N_c\,N_D\,L^d\,\R(m)}$, 
the expression of the continuum is recovered (except that the 
$\lambda$s are complex and not purely imaginary); in particular, if we write:
\be
\label{eq:fac}
\Z(\theta,m)\:=\:e^{\f{1}{2}N_c\,N_D\,L^d\,\R(m)}\,\Z'(\theta,m)\;\;;
\ee
we see that in the limit (\ref{eq:lim}) $\Z'(\theta,m)$ 
is invariant under the symmetry 
(applying in the continuum for $Z(\theta,m)$):
\bea
\label{eq:sym}
m\:&\rightarrow&\: e^{i\phi}\,m \nonumber \\
\theta\:&\rightarrow&\: \theta-\phi\;\;,
\eea
which implies 
\be
\label{eq:par}
\Z'(\theta,m)\: =\: \Z''(me^{i\theta})\;\;.
\ee 
If the theory has a mass gap non--vanishing for 
$m\rightarrow 0$ (which is true in the one flavor case), 
in the infinite volume limit we can assume
\be
\Z(\theta,m)\:=\:\exp\left\{-L^d\epsilon_0(\theta,m)\right\}\;\;,
\ee
where $\epsilon_0(m,\theta)$ is the lattice analogous of the 
vacuum energy density.
The corrections to this relation are exponentially small,
$O(e^{-m_0L})$, where $m_0$ corresponds to the mass--gap of 
the theory \cite{LeSm92}.

Using the factorization property (\ref{eq:fac})--(\ref{eq:par})
we parametrize $\epsilon_0(\theta,m)$ as
\be
\label{eq:part}
\epsilon_0(\theta,m)\:=\:C\,-\,\Sigma\,\R(m\,e^{i\theta})\,-\,
\f{1}{2}N_c\,N_D\,\R(m)\,+\,O(m^2)\;\;;
\ee
the last but one term is an ultraviolet divergent
and topology--independent contribution to the vacuum energy density 
$\sim 1/a^{D-1}$. It is a lattice artifact (absent in the 
continuum \cite{LeSm92}) appearing because of the explicit breaking 
of the chirality; it must be subtracted in order 
to get the correct continuum limit. The parameter $\Sigma$ is expected to
scale as a physical quantity of dimension $D-1$ and gives a lattice 
definition of the fermion condensate in the infinite volume limit.
From (\ref{eq:part}) it follows:
\be
\label{eq:subco}
\Sigma\:=\:-\,\langle\bar{\psi}_x\psi_x\rangle_{m=0,\,L\rightarrow\infty}\,
-\,\f{1}{2}N_c\,N_D\;\;;
\ee
observe that in the case of absence of massless excitations
the two limits $L\rightarrow\infty$ and $m\rightarrow 0$ 
can be interchanged. 

In the framework of Ginsparg--Wilson fermions a subtracted fermion condensate
can be defined \cite{Ha98a} (see also the discussion in \cite{Ne98d}), 
which for $N_f>1$ represents an order parameter 
for spontaneous breaking of the $SU(N_f)_A$ symmetry:
\be
\subcond\:=\:\langle{\rm tr}^{DFC}(-\,\D_{x,x}^{-1}\,+\,R_{x,x})
\rangle_{\rm gauge}\;\;,
\ee
where ${\rm tr}^{DFC}$ is a trace over Dirac, flavor and color indices
and $\langle\cdots\rangle_{\rm gauge}$ denotes 
the gauge average (including the fermion determinant);
in our case, where $R_{x,x'}=(1/2)\,\delta_{x,x'}$ 
and $N_f=1$, we argue from (\ref{eq:subco})
\be
\Sigma\:=\:-\,\subcond\;\;.
\ee

Using  (\ref{eq:part}) the calculation
of $\Z_{\nu}(m)$ is readily worked out:
\be
\label{eq:zeta}
\Z_{\nu}(m)\:=\:\int\,d\theta\,e^{-i\nu\theta}\,Z(\theta,m)\:=\:
\left(\f{m}{|m|}\right)^{\nu}\, e^{\f{1}{2}N_c\,N_D\,L^D\,\R(m)}\,
I_{\nu}(L^d\Sigma|m|)\;\;.
\ee

\section{Leutwyler--Smilga sum rules}

Now we exploit Leutwyler and Smilga's idea to obtain the wanted sum rules;
we write (from now on we take $m$ real and positive and $\theta=0$): 
\be
\label{eq:trick}
\Z_{\nu}(m)\:=\:m^{\nu}\,\int[dU^{(\nu)}]\,e^{-\beta S_G}\,
\left({\prod_i}'\lambda_i\right)\,\left[{\prod_i}'\left(1+\f{m}{\lambda_i}
\right)\right]\;\;,
\ee
where the primed product is intended over non--zero modes.
We rewrite 
\be
\label{eq:prod2}
{\prod_i}'(1+\f{m}{\lambda_i})\:=\:(1+\f{m}{2})^{N_c\,N_D\,L^d-\nu}\,
{\prod_i}''\left[1+\left(\f{m}{1+\f{m}{2}}\right)^2\,\f{1}
{\tilde{\lambda}^2}\right]
\ee
where we have introduced the new real variable:
\be
\tilde{\lambda}(\lambda)\:=\:\f{-i\,\lambda}{1-\f{1}{2}\lambda}\;\;;
\ee
geometrically, $\tilde{\lambda}(\lambda)$ is obtained from $\lambda$
by the stereographic projection of the unit circle centered in (1,0)
(where $\lambda$ lives) onto the imaginary axis; clearly
$\tilde{\lambda}(\bar{\lambda})=-\tilde{\lambda}(\lambda)$.

In the limit (\ref{eq:lim}) the r.h.s. of (\ref{eq:prod2}) may be replaced by
\be
\label{eq:prod}
e^{\f{1}{2}N_c\,N_D\,L^d\,m}\, 
{\prod_i}''\left(1+\f{m^2}{\tilde{\lambda}^2}\right)\;\;,
\ee
again an expression analogous to the continuum one apart from the 
multiplicative factor $e^{\f{1}{2}N_c\,N_D\,L^d\,m}$.
Inserting (\ref{eq:prod}) in the r.h.s. of (\ref{eq:trick}),
we come to the relation, exact in the limit (\ref{eq:lim}): 
\be
\langle\langle\,{\prod_i}''\left(1+\f{m^2}{\tilde{\lambda}^2}\right)\,
\rangle\rangle_{\nu}\:=\:e^{-\,\f{1}{2}N_c\,N_D\,L^d\,m}\,
\f{m^{-\nu}\Z_{\nu}(m)}{\lim_{m\rightarrow 0}
\left(m^{-\nu}\Z_{\nu}(m)\right)}
\ee
where $\langle\langle\cdots\rangle\rangle_{\nu}$
is the average over gauge configurations with associated index $\nu$
in the massless case, the fermion determinant 
being replaced by the product of non--zero eigenvalues
of the Dirac operator. 
Using the explicit representation (\ref{eq:zeta}) 
and
\be
I_{\nu}(x)\:\simeq\:\f{1}{|\nu|!}\,\left(\f{x}{2}\right)^{|\nu|}\;\;,\;\;\;\;x\ll 1\;\;,
\ee
we obtain (for any sign of $\nu$)
\be
\label{eq:master}
\langle\langle\,{\prod_i}''\left(1+\f{m^2}{\tilde{\lambda}^2}\right)
\,\rangle\rangle_{\nu}\:=\:|\nu|!\left(\f{2}{x}\right)^{|\nu|}\, I_{\nu}(x)\;\;,
\ee
where $x=L^d\Sigma m$.
So we recover formally the same result of the continuum, from which
Leutwyler--Smilga sum rules originate, with the only difference
that in the sums the original eigenvalue $\lambda$ is replaced 
by the projected one $\tilde{\lambda}$. \footnote{Or any other definition 
$\tilde{\lambda}'$ such that 
$\tilde{\lambda}'=\tilde{\lambda}(1+O(\tilde{\lambda}))$}
For example  
from (\ref{eq:master}) it follows \cite{LeSm92}
\be
\lim_{L\rightarrow\infty}\,\f{1}{(L^d\,\Sigma)^2}\,\langle\langle\,{\sum_i}''\,\f{1}{\tilde{\lambda_i}^2}\,\rangle\rangle_{\nu}\:=\:
\f{1}{4(\nu+1)}\;\;.
\ee

{\bf Acknowledgment:}
We are grateful to C.~B. Lang and K. Splittorff 
for stimulating discussions on the subject. 
Support by Fonds zur F\"orderung der Wissenschaftlichen Forschung 
in \"Osterreich, Project P11502-PHY is gratefully acknowledged.

\end{document}